\documentstyle[12pt,epsf]{article}
\newcommand{\bi}{\bibitem}
\newcommand{\be}{\begin{eqnarray}}
\newcommand{\ee}{\end{eqnarray}}
\newcommand{\nn}{\nonumber}
\catcode`\@=11
\def\lsim{\mathrel{\mathpalette\@versim<}}
\def\gsim{\mathrel{\mathpalette\@versim>}}
\def\@versim#1#2{\vcenter{\offinterlineskip
\ialign{$\m@th#1\hfil##\hfil$\crcr#2\crcr\sim\crcr } }}
\catcode`\@=12

\begin{document}

\begin{flushright}
KANAZAWA-00-11\\
November 2000
\end{flushright}

\begin{center}
{\Large\bf 
Extra dimensions prefer large $\tan\beta$.}
\end{center} 

\vspace{1cm}
\begin{center} {\sc Jisuke Kubo} and {\sc Masanori Nunami} 
\end{center}

\begin{center}
{\em 
Institute for Theoretical Physics, 
Kanazawa  University, 
Kanazawa 920-1192, Japan
}
\end{center}

\vspace{1cm}
\begin{center}
{\sc\large Abstract}
\end{center}

\noindent
Assuming that the recent result obtained from the 
Monte Carlo simulations on
the nonperturbative existence of
the pure $SU(2)$ Yang-Mills theory in five dimensions
can be  applied to a more general class of
higher-dimensional gauge theories , we derive the conditions imposed by
the nontriviality requirement on the theories.
We find that 
the supersymmetric grand unified theories with extra dimensions 
prefer a large value 
($\gsim 2$) of $\tan\beta$ of the minimal supersymmetric
standard model, in accord with
today's possible observation of  the Higgs particle at LEP2.

\vspace{1cm}

\newpage
\section{Introduction}
In recent years a variety of theories
having large extra space-time dimensions have been 
considered \cite{antoniadis1,witten1,lykken1,arkani1,antoniadis2,dienes1,randall1}.
It has been found that certain  
theoretical problems
such as  the unification of the fundamental forces and
the hierarchy problem 
may be solved by introducing 
 large  extra dimensions (see Ref. \cite{antoniadis3,dine} for review).
So far it is only a theoretical speculation that we live 
in more than four space-time dimensions, and 
experimental indications for the existence of 
extra  dimensions are currently searched \cite{antoniadis3,dienes2}.

It is widely believed that
any interacting gauge theory in more than four dimensions,
being perturbatively unrenormalizable, 
is a cutoff theory, and that  for a certain range of energy scale
it can be an effective theory of a more fundamental theory such as string theory. 
Is it possible to control the quantum corrections
in gauge theories in more than four dimensions?
Is it  ensured that the massive
Kaluza-Klein excitations below the compactification scale
really decouple so that its effective theory below that scale
becomes  a four-dimensional renormalizable theory?
How can we answer these questions?
The investigation of the nonperturbative existence of
gauge  theories in higher dimensions is, therefore,  not only an academic
problem, but also a fundamental problem if the fundamental theory
of particle physics is formulated in more than four 
dimensions. Recently, 
 the pure $SU(2)$ lattice gauge
theory in five dimensions has been investigated \cite{ejiri1}, 
where the extra
dimension is assumed to be compactified on
a circle with the radius of $R$. 
It has been found there that
the scaling behavior of  the Creutz
ratio measured in the
four-dimensional subspace  indicates that 
the compactified theory with a nonvanishing string tension
can exist nonperturbatively.
That is, the investigation indicates that the theory
is a cutoff-free theory.
Interestingly, this observation 
is consistent with the existence of
the nontrivial ultraviolet
fixed point that can be found 
analytically in the $\epsilon$-expansion method \cite{peskin1}.

It is well conceivable 
that not only the pure $SU(2)$ Yang-Mills theory in five dimensions 
can exist nonperturbatively, but also a more general class of 
higher-dimensional Yang-Mills theories containing
bosonic and fermionic matter fields in various representations.
Unfortunately, because of the lack of computer power,
these investigations based on  lattice gauge theories
are limited, and phenomenologically interesting
higher-dimensional unified gauge  models will not be accessible 
within the framework of lattice gauge theory in near
future. 
We therefore assume
that the  fact \cite{ejiri1} that 
the lattice $\beta$ function of the
gauge coupling can be well approximated by the one-loop
form can be extended to other cases, and that
the existence of an ultraviolet fixed point can be
investigated on the basis of the one-loop $\beta$ functions.
In doing so we would like to derive the conditions imposed by
the requirement of the
nontriviality of the higher-dimensional 
unified gauge theories. We expect phenomenological
consequences from this requirement, 
as the upper bound of the Higgs mass  of
the standard model (SM) can be obtained from the  nontriviality 
requirement of the model \cite{ms-higgs}.

In Sect. II we start by summarizing  the results from the Monte Carlo simulations
in the pure $SU(2)$ Yang-Mills theory in five dimensions to 
make clear our assumptions about the nontriviality
of a more general class of higher-dimensional  Yang-Mills theories.
In Sect III we will derive the conditions 
for  a supersymmetric grand unified theory (SUSY GUT) to   be nontrivial, and
apply in Sect. IV
this result to a concrete model based on the gauge group
$SU(5)$ in $4+\delta$ dimensions.
We will find that the
nonperturbative existence of the model requires a large value 
($\gsim 2$) of $\tan\beta$ of the minimal supersymmetric 
standard model (MSSM), and that this is a general feature of
SUSY GUTs with extra dimensions, suggesting that
today's possible observation of the 
Higgs particle with the mass $\sim 115$ GeV at LEP2 \cite{lep2} 
could be an indication for the existence of extra dimensions.

\section{The lattice result and its generalization}

As mentioned in Introduction,  the pure $SU(2)$ lattice gauge
theory in five dimensions has been investigated in Ref. \cite{ejiri1}, 
where an extra dimension is  compactified on a circle with the radius  $R$. 
There, anisotropic lattices \cite{burgers,klassen,scheideler,ejiri2}
have been insensitively used to extract
maximally the compactification effects, and it has been
observed that 
the  first order phase transition  which
exists in the uncompactified case \cite{creutz1,okamoto1,nishimura1}
changes its nature at a certain 
compactification radius, and becomes of second order.
Moreover, it has become possible \cite{ejiri1}, through compactification,
to compute the $\beta$ function of the gauge coupling, which
in turn shows its power-law running, and has been found that the
observed power law behavior fits well to the one-loop form suggested 
in  perturbation theory \cite{veneziano,dienes1}. 
Of course, the nonperturbative existence of a theory or the existence
of an nontrivial ultraviolet fixed point in the theory should not
depend on whether some extra dimensions are compactified or not.
We therefore believe that
the compactification that has been assumed in 
the above mentioned investigations based on the
lattice regularization is only technically 
indispensable, and that the theory
exists nonperturbatively whether the extra dimension is  compactified or not.

It is natural to assume that
not only the pure $SU(2)$ Yang-Mills theory in five dimensions 
can exist nonperturbatively, but also a wide class of 
higher-dimensional Yang-Mills theories.
Thought out this paper we  assume
that the fact that
the lattice $\beta$ function of the
gauge coupling in the pure $SU(2)$ Yang-Mills theory in five dimensions 
can be well approximated by its one-loop
form can be extended to other higher-dimensional Yang-Mills
theories, and that
the existence of an ultraviolet fixed point can be
investigated on the basis of the one-loop $\beta$ functions
in these theories.

Let us explain more in detail our assumption in the case of the 
pure $SU(N_C)$ Yang-Mills theory in $D$ dimensions,
where we assume that $\delta=D-4$ dimensions are compactified
on a circle with the radius of $R$. 
Let $g_{DYM}$ be the  gauge coupling of the theory. Then
the dimensionless, four-dimensional gauge coupling
of the compactified theory
is defined as
\be
g &=& (2\pi R)^{-\delta/2}g_{DYM}~.
\ee
The compactified theory has an infinite tower of the massive
Kaluza-Klein states (at least at the classical level).
 We think of integrating out these
massive modes down to the cutoff energy $\Lambda$
and define an effective theory at $\Lambda$.
So, at the quantum level, the dimensionless gauge coupling $g$
is the effective gauge coupling
and is a function of $\Lambda$.
The $\beta$ function
of $g$, 
\be
\Lambda\frac{d g}{d \Lambda} &=&\beta_g^{(1)}+\cdots~,
\ee
takes in the one-loop order  the form \cite{dienes1}
\be
\beta_g^{(1)} &=&-\frac{1}{16\pi^2}~b_0~
(R\Lambda)^{\delta}~X_\delta~g^3~,~b_0=\frac{22-1}{6}N_C~.
\ee
The coefficient $X_\delta$ is 
a regularization-dependent constant \cite{veneziano,kobayashi1,kakushadze2}, 
and
in the proper time regularization scheme
employed in Ref. \cite{dienes1} it is given by
\be
X_{\delta}=\frac{\pi^{\delta/2}}{\Gamma(1+\delta/2)} ~.
\label{x-delta}
\ee
We have added to $b_0$ the contribution ($-(1/6) N_C$ in $b_0$) 
coming from the scalar in the adjoint representation.
The power law $(R\Lambda)^{\delta}$ expresses the fact that
the lager the cutoff $\Lambda$ 
is the more states are circulating in a loop.
This power-like growing of the number of states can
be absorbed into a redefinition of the coupling
\be
\hat{g} &=& (2\pi R\Lambda)^{\delta/2}~g
=\Lambda^{\delta/2}~g_{DYM}~,
\label{g-hat}
\ee
whose   $\beta$ function becomes
\be
\Lambda\frac{d \hat{g}}{d \Lambda} &=&\hat{\beta}_g^{(1)}+\cdots
=\frac{\delta}{2}~\hat{g}
-\frac{1}{16\pi^2}~b_0~
~\frac{X_\delta~}{(2\pi)^\delta}~\hat{g}^3+\cdots.
\label{g-hat-beta}
\ee
We see now that the $\beta$ function of $\hat{g}$ 
can have a nontrivial  ultraviolet fixed point at \footnote{This is 
the critical value in investigating whether or not
the dynamical electroweak symmetry breaking by the top condensation in
higher dimensions \cite{arkani2} can occur \cite{hashimoto}.}
\be
(\hat{g}^*)^2 &=&\frac{\delta}{2}~\frac{16 \pi^2}{b_0}
\frac{(2\pi)^\delta}{X_\delta}~.
\label{g-hat-star}
\ee
The data obtained from the  Monte Carlo  simulations 
for the pure $SU(2)$ gauge theory in five dimensions \cite{ejiri1} indicate
that the ultraviolet fixed point (\ref{g-hat-star}) 
in this case is  indeed a real
one. Eq. (\ref{g-hat-beta})  
suggests that the redefined, dimensionless
gauge coupling  $\hat{g}$, rather than $g$,  can be 
regarded as the effective expansion parameter.
Our central assumption 
is thus that one can decide on the
nonperturbative existence of a higher-dimensional Yang-Mills theory
 from the investigation  of
the ultraviolet fixed points in the space of the effective
expansion parameters at the one-loop level.

\begin{figure}
\epsfxsize= 12 cm
\centerline{\epsffile{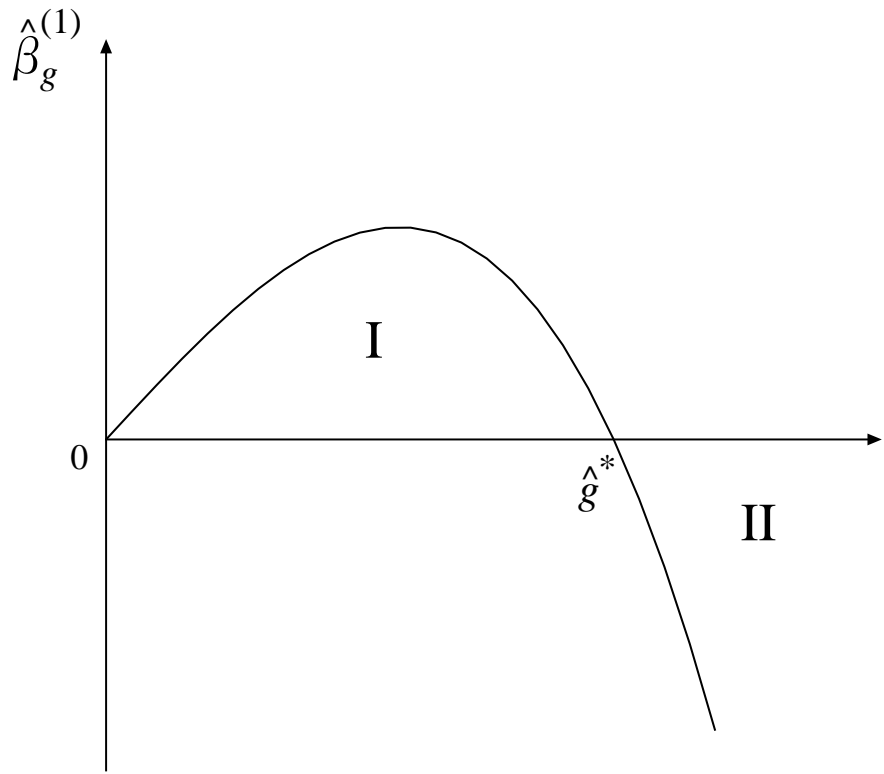}}
\caption{The generic form of $\hat{\beta}_{g}^{(1)}$.}
        \label{fig:1}
        \end{figure}
\begin{figure}
\epsfxsize= 12 cm
\centerline{\epsffile{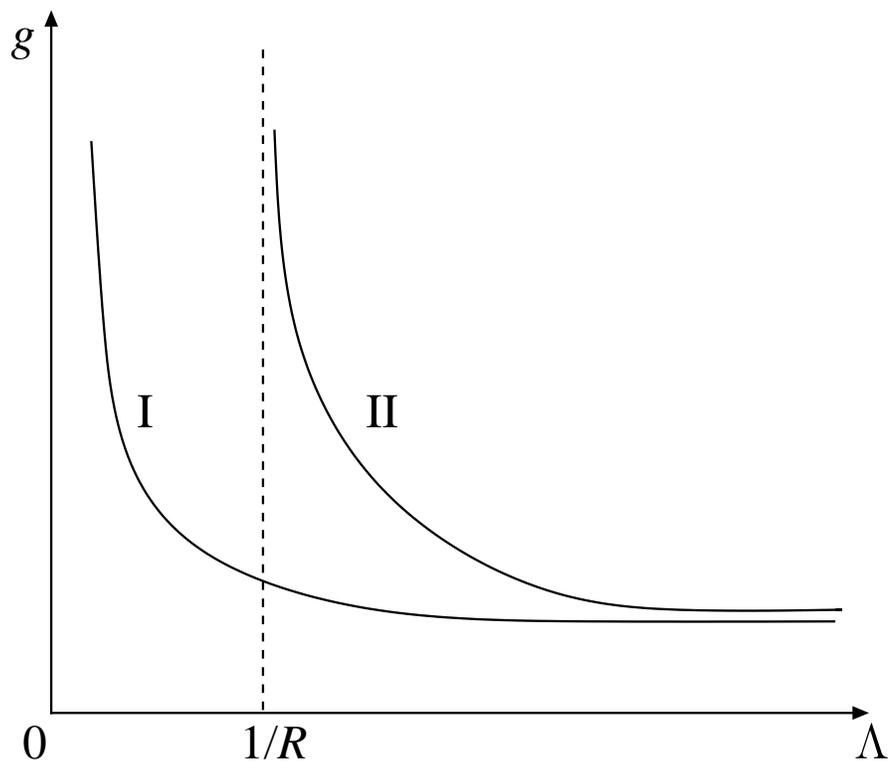}}
\caption{ The evolution of the gauge coupling
in two phases. Phase I is the decoupling phase, while
phase II is the strongly interacting phase.}
        \label{fig:2}
        \end{figure}

The generic form of the $\beta$ function $\hat{\beta}^{(1)}_g$
[see Eq. (\ref{g-hat-beta})]
is shown in Fig. 1, in which two phases are indicated by I
and II. The renormalization group (RG) 
flow of the gauge coupling $g$ in two phases are different, as shown 
in Fig. 2. As the energy scale $\Lambda$ decreases from a higher value, 
the flow of the
phase II develops into a ``Landau'' pole near the compactification scale $\sim 1/R$, 
while the coupling in phase I has no such singularity near $\sim 1/R$.
That is, the theory in phase II will become strongly interacting
near $\sim 1/R$, and it will be unlikely that the massive
Kaluza-Klein excitations (which seem to exist at the classical level)
decouple 
\footnote{ Presumably, the notion of
the massive Kaluza-Klein excitations is not a good one in phase II.
Moreover, it is unclear that the low-energy effective 
theory in phase II is 
a gauge theory.}.
Only if the theory is in phase I there will be a chance
for the massive Kaluza-Klein excitations to decouple and
hence to have a renormalizable, low-energy 
effective theory below  $\sim 1/R$.
We regard this as a constraint on the gauge coupling.
So the theory should be
in phase I; in the decoupling phase.

\section{Nontriviality of SUSY GUTs with extra dimensions}
We assume 
that  $\delta=D-4$ dimensions are compactified
on an orbifold $S^1/Z_2$ of a fixed radius $R$.
We denote the $D$-dimensional coordinates by $z_M
~(M=1,\dots,D)$,
while  the four-dimensional ones by $x_{\mu}~(\mu=1,\dots,4)$
and  the $\delta$-dimensional ones by 
$y_{a}~(a=1,\dots,\delta)$.
A generic  field $\phi(z)$, bosonic or fermionic,  satisfying the periodic 
 boundary condition 
\be
\phi(x,y) &=&\phi(x,y+2\pi R)~
\ee
with the parity property under $y_a \to -y_a$
\be
\phi &=& \phi~|_{y_a \to -y_a}~~\mbox{for}~~
a \in E_{+} ~~\mbox{and}~~
\phi = -\phi~|_{y_a \to -y_a}~~\mbox{for}~~
a \in  E_{-}
\ee
can be expanded as
\be
\phi(x,y) &=&
 ~\sum_{{\bf n}=0}^{\infty}~
 ~\sum_{{\bf m}=1}^{\infty} ~\phi_{{\bf n},{\bf m}}(x)
~\Pi_{a \in E_+}
~\cos ( n_a y_a/R)~ \Pi_{b \in E_-}~ \sin ( m_b y_b/R)~,
\ee
where we have divided $E=\{a=1,\dots,\delta \}$ into
$E_+$ and $E_-$ corresponding to the parity property
of $\phi$.  The coefficients $\phi_{{\bf n},{\bf m}}(x)$
exhibit the Kaluza-Klein tower, and
$\phi_{{\bf 0},{\bf 0}}(x)$
is  the zero mode, which is absent if $\phi$ 
has an odd parity. The Kaluza-Klein modes other than the zero mode are
massive $\sim O(1/R)$ in four dimensions. Since we consider GUTs, a certain
set of the  zero modes also become massive after a spontaneous symmetry
breaking of  the unifying gauge group $G$. Their masses are of the order of
the spontaneous symmetry breaking or of the GUT scale $M_{G}$.
The presence of the fields that exist only at a lower-dimensional
boundary, the boundary fields,  is allowed in the case of the orbifold
compactification. Here we restrict ourselves  only to
the boundary fields that are located at our four-dimensional Minkowski
space. They have no Kaluza-Klein massive partners, and
they  count among the  zero modes.

Our main assumptions in considering SUSY GUTs are
that (i) in the zero
mode sector of the Kaluza-Klein tower,
softly broken, four-dimensional $N=1$ supersymmetry is realized 
and (ii)
the massive Kaluza-Klein modes form $N=2$ supermultiplets.
The first assumption can be simply satisfied
thanks to the orbifold compactification, and
the second one can also be easily satisfied
because a simple supersymmetry in higher dimensions
always contains more than one supersymmetry in four dimensions. 
Correspondingly, the matter supermultiplets 
of the zero mode sector are
$N=1$ chiral supermultiplets,
\be
\Phi_I &=&(\phi_I~,~\psi_I)~,
\ee
where $\phi_I~(\psi_I)$ 
is the scalar (fermionic) component, and
$I$ stands for color and flavor.
The most general (cubic) form of the
Yukawa term of the zero mode sector at
the four-dimensional boundary takes the form
\be
S_0^Y &=& \int d^4 x~\frac{1}{2}~
\sum_{I,J,K} Y^{IJK}~\psi_I \psi_J \phi_K~+\mbox{h.c.},
\ee
where the  Yukawa couplings  $Y^{IJK}$ are assumed to be
completely symmetric in the indices.
Although we have to add a set of certain terms to the above
action $S_0^Y$ to make the boundary  theory supersymmetric and
gauge invariant, the complete space of the 
dimensionless couplings
of the boundary theory, by virtue of 
$N=1$ supersymmetry, is spanned by the gauge coupling $g$
and the Yukawa couplings $Y^{IJK}$:
That is,  no additional dimensionless
couplings are present.

If the contributions of the massive Kaluza-Klein
modes to the RG functions 
($\beta$ functions and anomalous dimensions $\gamma$) 
are suppressed,  we have the well-known 
four-dimensional formulae \cite{jones1}:
\be
\beta^{(1)}_{g} &=&
\frac{g^3}{16\pi^2}\,[\,\sum_{a}\,l(R_{a})-3\,C_{2}(G)\,]~
\ee
at one-loop, where $l(R_{a})$ is the Dynkin index of the 
representation $R_{a}$ and $C_{2}(G)$
 is the
quadratic Casimir of the adjoint representation of the
gauge group $G$. The $\beta$ functions of
$Y^{IJK}$ are related to the
anomalous dimensions $\gamma_{I}^{(1)\,J} $ as \cite{jones1}
\be
\beta_{Y}^{(1)IJK} &= & \sum_{P}~Y^{IJP}\,
\,\gamma_{P}^{(1)\,K} +(K
\leftrightarrow I) +(K\leftrightarrow J)~,\\
\gamma_{I}^{(1)\,J} &= &\frac{1}{16\pi^2}~[\,
\frac{1}{2}\sum_{P,Q}~ Y_{IPQ} Y^{JPQ}
-2\delta^{J}_{I}\, g^2\,C_2(I)\,]~,
\ee
where $C_{2}(I)$ is the quadratic Casimir of the representation
$R_{I}$, and $Y^{IJK}=(Y_{IJK})^{*}$.

The Kaluza-Klein tower modifies  the RG functions
to the form that describes the power-law behavior of the couplings.
The inclusion of the contribution of 
the massive Kaluza-Klein modes to the RG functions
is straightforward, because they form $N=2$ supermultiplets
by assumption and we may use the nonrenormalization theorem for
$N=2$ supersymmetry \cite{howe}. 
Among the zero modes, there are those
that have no massive partner modes, and they do not contribute
to  the power law behavior of the couplings.
Therefore, their contributions to
 the RG functions in the $\Lambda \to \infty$ limit are much smaller
compared with those coming from the infinite tower of the massive modes, i.e.,
\be
\beta_{g}^{(1)} &\simeq&
\frac{g^3}{16\pi^2}\,[\,\sum_{a'}\,l(R_{a'})-2\,C_{2}(G)\,]
~X_\delta ~(\Lambda R)^{\delta}~,
\label{beta-g-x}\\
\beta_{Y}^{(1)IJK} &\simeq & \frac{1}{16\pi^2}\,[~\frac{1}{2}~
\sum_{P,L,M}{}'~Y^{IJP}Y_{PLM}Y^{KLM}
-2 \sum_{P}{}' ~ Y^{IJ P}\, \delta_P^K ~g^2\,C_2(K)\nn\\
& & +(K\leftrightarrow I) +(K\leftrightarrow J)~]
~X_\delta ~(\Lambda R)^{\delta}~,
\label{beta-y-x}
\ee
where $ X_\delta$ is given in Eq. (\ref{x-delta}).
Here $\sum_{a'}$ denotes the sum over $N=2$ hypermultiplets,
and  $\sum {}'$ denotes the sum in which only  the possibilities
that contribute to the power law behavior are included.
In deriving the $\beta$ functions
(\ref{beta-g-x}) and (\ref{beta-y-x}), we have used the fact that 
the contributions of each excited
Kaluza-Klein state to the anomalous dimension has the same form as a
massless mode contribution \cite{dienes1}.

Now according to the discussion in the previous section, we go
over to the effective expansion parameters:
As for the gauge coupling, it is defined in Eq. (\ref{g-hat}),
and similarly we can find  them for the Yukawa couplings. 
It is however more convenient to work with
\be
\tilde{Y}^{IJK} &=& \frac{Y^{IJK}}{g}=
 \frac{Y^{IJK}(2\pi R\Lambda)^{\delta/2}}{g(2\pi R\Lambda)^{\delta/2}}
= \frac{\hat{Y}^{IJK}}{\hat{g}}~,
\label{y-tilde}
\ee
which yields the following  system of the  $\beta$ functions
at large $\Lambda$:
\be
\hat{\beta}_{g} &=&\Lambda\frac{d \hat{g}}{d \Lambda} =
\frac{\delta}{2}\hat{g}
-\frac{\hat{g}^3}{16\pi^2}\,[\,2\,C_{2}(G)\sum_{a'}-\,l(R_{a'})\,]
~\frac{X_\delta}{(2\pi)^{\delta}}+\cdots~,
\label{beta-g-hat}\\
\tilde{\beta}_{Y}^{IJK} &=& 
\Lambda\frac{d \tilde{Y}^{IJK}}{d \Lambda}=
\hat{g}^2 ~[~-\tilde{Y}^{IJ K} S(IJK)+(~\frac{1}{2}~
\sum_{P,L,M}{}'~\tilde{Y}^{IJP}\tilde{Y}_{PLM}\tilde{Y}^{KLM}~)\nn\\
& &+(K\leftrightarrow I) +(K\leftrightarrow J) 
~]~
\frac{X_\delta}{(2\pi)^{\delta}}+\cdots~,
\label{beta-y-tilde}
\ee
where $\cdots$ stands for higher order contributions, and
\be
\tilde{Y}^{IJ K} S(IJK)
& =&\tilde{Y}^{IJ K}
(\,\sum_{a'}\,l(R_{a'})-2\,C_{2}(G)\,)\nn\\
& &+[~2 \sum_{P}{}' ~ \tilde{Y}^{IJ P}\, \delta_P^K \,C_2(K)~]
+[K\leftrightarrow I] +[K\leftrightarrow J]~.
\label{s-ijk}
\ee
Note that the sum in Eq. (\ref{s-ijk})  (though it is proportional
to  $\tilde{Y}^{IJ K}$) is not equal to
$2 \tilde{Y}^{IJ K}(C_2(I)+C_2(J)+C_2(K))$,
because the sum $\sum_{P}{}'$ is taken over
 only   the possibilities
that contribute to the power law behavior.

From the $\beta$ functions (\ref{beta-g-hat}) and 
(\ref{beta-y-tilde}) we see that 
\be
\tilde{Y}^{IJK} &=& 0 
\label{trivial-fp}
\ee
is an ultraviolet stable fixed point, if 
\be
S(IJK) & > & 0 ~~\mbox{for all}~~I, J, K
~~\mbox{and}~~2\,C_{2}(G)-\sum_{a'}\,l(R_{a'}) > 0~
\label{cond1}
\ee
are satisfied.
According to our assumption,
the theory exists nonperturbatively if the conditions
(\ref{cond1}) are satisfied.
If $S(IJK)> 0$, on the other hand, there will be a certain set
of infrared fixed points. 
That is, the stable manifold (the set of points
in the space of  $\tilde{Y}^{IJK}$ that can be initial points
of a RG flow approaching the ultraviolet fixed 
point (\ref{trivial-fp}) )
must be a subspace of the 
space of $\tilde{Y}^{IJK}$. Therefore,
the requirement of the nonperturbative existence
implies that the Yukawa couplings  at the GUT scale $M_G$
should be  in the stable manifold.
If all the Yukawa couplings are small compared with
the unified gauge coupling, this condition can be easily
satisfied. If however some of the Yukawa couplings, e.g.
the top quark Yukawa coupling, are comparable with
the unified gauge coupling in the magnitude, this condition can be a severe condition.
The situation depends on the model considered of course.
In the next section we consider a concrete gauge model based
on the gauge group $SU(5)$ and discuss how the 
requirements coming from the nontriviality can be satisfied.

\section{Application to the minimal
SUSY $SU(5)$ GUT}

\subsection{The model and its nontriviality}
The three
generations of quarks and leptons   are accommodated 
 \footnote{We use the four-dimensional
language  for supersymmetry.} by 
six chiral $N=1$ 
 superfields 
$\Psi^{i}({\bf 10})$ and $\Phi^{i}(\overline{\bf 5})$,
where $i$ runs over the three generations.
The superfield $\Sigma({\bf 24})$ is used to break $SU(5)$ down to $SU(3)_{\rm C}
\times SU(2)_{\rm L} \times U(1)_{\rm Y}$,  and
$H({\bf 5})$ and $\overline{H}({\overline{\bf 5}})$
(which form an $N=2$ hypermultiplet)
are
two Higgs superfields appropriate for electroweak 
symmetry breaking.
We assume that the matter superfields
$\Psi^{i}({\bf 10})$ and $\Phi^{i}(\overline{\bf 5})$
are boundary fields so that they have no Kaluza-Klein
excitations, and that  the $4+\delta$-dimensional
bulk theory is an
$N=2$ 
supersymmetric Yang-Mills theory
based on $G=SU(5)$ that contains a hypermultiplet in the fundamental
representation of $G$.
The cubic part of the boundary superpotential is given by
\be
W &=& \sum_{i,j}^{3}\frac{G_{U}^{ij}}{4}\,
\epsilon^{\alpha\beta\gamma\delta\tau}\,
\Psi^{(i)}_{\alpha\beta}\Psi^{(j)}_{\gamma\delta}H_{\tau}+
\sum_{i,j}^{3}\sqrt{2}\, G_D^{ij}\,\Phi^{(i) \alpha}
\Psi^{(j)}_{\alpha\beta}\overline{H}^{\beta}+
\frac{g_{\lambda}}{3}\,\Sigma_{\alpha}^{\beta}
\Sigma_{\beta}^{\gamma}\Sigma_{\gamma}^{\alpha}~,
\ee
where $\alpha,\beta,\ldots$ are the $SU(5)$
indices, and $G_{U}^{ij}$ and $G_{D}^{ij}$
are the Yukawa couplings. The
$\overline{H}\Sigma H$ term is a part of the $N=2$  
gauge interaction and  belongs to the bulk action.
To make the theory realistic, we have to have
the correct pattern of 
spontaneous symmetry breaking of gauge symmetries,
soft-supersymmetry-breaking (SSB) terms, 
and neutrino masses and
their mixing.
Note that only operators with dimensions less than four
are responsible to satisfy these phenomenologically
important requirements. 
Since however the contribution of these low-dimensional
operators to  the high
energy behavior of the theory decreases with an increasing
energy scale $\Lambda$, we  ignore them
in the following discussions.

Given the model, it is straightforward to compute
the one-loop RG functions. We find that the one-loop anomalous dimensions of
the chiral superfields are given by
\be
16 \pi^2~\gamma_{10}&=& 
[\,-\frac{36}{5}\,g^2+
3\,G_U^{\dag} G_U+2\,G_D^{\dag} G_D\,]
~X_\delta ~(R\Lambda)^{\delta}~, \\
16 \pi^2~\gamma_{\bar{5}}&=&
[\,-\frac{24}{5}\,g^2+
4\,G_D^{\dag} G_D\,]~
X_\delta ~(R\Lambda)^{\delta}~, \\
16 \pi^2~\gamma_{H} &=&
3\,\mbox{Tr}G_U G_U^{\dag}~,\\
~16 \pi^2~\gamma_{\bar{H}}&=&
4\, \mbox{Tr} G_D G_D^{\dag}~,\\
16 \pi^2~\gamma_{24} &=& [\,\frac{21}{5}
\,g^{2}_{\sigma}
-9\,g^2\,]~X_\delta ~(R\Lambda)^{\delta}~,
\ee
where $X_\delta$ is defined in Eq. (\ref{x-delta}).
Using Eqs. (\ref{beta-g-x}) and
(\ref{beta-y-x}) we then obtain the one-loop $\beta$ functions
of the Yukawa couplings, where we
use the fact that the
anomalous dimensions  of the Higgs supermultiplets 
 vanish thanks to $N=2$ supersymmetry.
Since we are interested
in the $\Lambda \to\infty$ limit, only leading contributions
in the limit,
\be
16 \pi^2~\beta_g &\simeq&
-9 g^3~X_\delta~ (R\Lambda)^{\delta}~,
\label{beta-ggut}\\
~16 \pi^2~\beta_U &\simeq& 
G_U~[\,-\frac{72}{5}\,g^2+
6\,G_U^{\dag} G_U+4\,G_D^{\dag} G_D\,]
~X_\delta ~(R\Lambda)^{\delta}~, \\
~16 \pi^2~\beta_D &\simeq& G_D~
[\,-12\,g^2+3\,G_U^{\dag} G_U+
6\,G_D^{\dag} G_D\,]~
X_\delta ~(R\Lambda)^{\delta}~,\\
16 \pi^2~\beta_{\sigma} &\simeq& g_\sigma~[\,-27\,g^2
+\frac{63}{5}\,g^{2}_{\sigma}
\,]~X_\delta ~(R\Lambda)^{\delta}~, 
\ee
should be considered.
According to the discussion of the previous section,
we now go over to the tide-couplings [defined in Eq. (\ref{y-tilde})],
and find that the corresponding one-loop 
$\beta$ functions in the $\Lambda \to \infty$ limit can be written
as 
\be
16 \pi^2~\tilde{\beta}_U/\hat{g}^2 &=& 
\tilde{G}_U~[\,-\frac{27}{5}\,+
6\,\tilde{G}_U^{\dag} \tilde{G}_U+4\,\tilde{G}_D^{\dag} 
\tilde G_D\,]~\frac{X_\delta}{(2\pi)^{\delta}}~,\\
16 \pi^2~\tilde{\beta}_D/\hat{g}^2 &=& \tilde{G}_D ~
[\,-3\,+3\,\tilde{G}_U^{\dag} \tilde{G}_U+
6\,\tilde{G}_D^{\dag} \tilde{G}_D\,]~\frac{X_\delta}{(2\pi)^{\delta}}~,\\
16 \pi^2~\tilde{\beta}_{\sigma}/\hat{g}^2 &= & \tilde{g}_\sigma~[\,-18+
\frac{63}{5}\,\tilde{g}^{2}_{\sigma}
\,]~\frac{X_\delta}{(2\pi)^{\delta}}~,
\ee
where $\hat{g}$ is defined
in Eq. (\ref{g-hat}).
Moreover, the phenomenological
requirements from the  mass of leptons and
quarks as well as from the proton decay \cite{hisano1,kubo1}
imply that the Yukawa couplings for the top and
bottom quarks, $G_{U}^{33}=G_t$ and $G_{D}^{33}=G_b$,
are the largest couplings compared with
the other Yukawa couplings, and therefore, to investigate
 approximately the high energy behavior of the theory, 
it is sufficient to 
consider the following set of the $\beta$ functions:
\be
16 \pi^2~\frac{\tilde{\beta}_t}{\hat{g}^2} &=& 
\tilde{G}_t~[\,-\frac{27}{5}\,+
6\,\tilde{G}_t^{2}+4\,\tilde{G}_b^{2} \,]~\frac{X_\delta}{(2\pi)^{\delta}}~,
\label{beta-t}\\
16 \pi^2~\frac{\tilde{\beta}_b}{\hat{g}^2} &=& \tilde{G}_b ~
[\,-3\,+3\,\tilde{G}_t^{2} +
6\,\tilde{G}_b^{2}\,]~\frac{X_\delta}{(2\pi)^{\delta}}~.
\label{beta-b}
\ee
We thus have arrived at a simple system defined by 
Eqs. (\ref{beta-t}) and  (\ref{beta-b}) that has four
fixed points 
\be
(~\tilde{G}_t^{* 2},\tilde{G}_b^{* 2} ~) &=&(~0,0 ~)~,~
(~\frac{17}{20},\frac{3}{40}~)~,
(~\frac{9}{10},0 ~)~,
(~0,\frac{1}{2}~)~
\label{fps}
\ee
in the two-dimensional space of couplings
$\tilde{G}_t^2$ and $\tilde{G}_b^2$, which are shown in Fig. 3.
As we have seen in the 
previous section, the origin $(~0,0 ~) $ is an ultraviolet-stable fixed point.
The point $(~17/20,3/40 ~) $ is an
infrared stable fixed point 
(the Pendleton-Ross fixed point 
\cite{pendleton}\footnote{The last three nontrivial
fixed points of the r.h.s. of (\ref{fps}) 
can be used to express  $\tilde{G}_t^2$ and $\tilde{G}_b^2$ 
in terms of the unified gauge coupling $g$ (reduction of couplings 
\cite{zimmermann}).}), while  for the
other  two points there exist
attractive as well as repulsive directions. We find that the direction
perpendicular to the $\tilde{G}_b^2$ axis is the 
infrared-attractive direction for
the fixed point $(~9/10,0 ~)$,
and similarly, 
the direction
perpendicular to the $\tilde{G}_t^2$ axis is the 
one for $(~0,1/2)$.
 In Fig. 3 we show some representative RG
flows, and as we can see from the figure, the stable manifold is 
a finite region in the space of $\tilde{G}_t^2$ and $\tilde{G}_b^2$. The
critical lines that go from 
the infrared stable point
$ (~17/20,3/40~)$ toward the end points $ (~9/10,0 ~)$ and $ (~0,1/2~)$
define the boundary of the stable manifold.
We emphasize that the result above is independent of the number
of the extra dimensions $\delta$ and the scale $\Lambda$.
\begin{figure}
\epsfxsize= 12 cm
\centerline{\epsffile{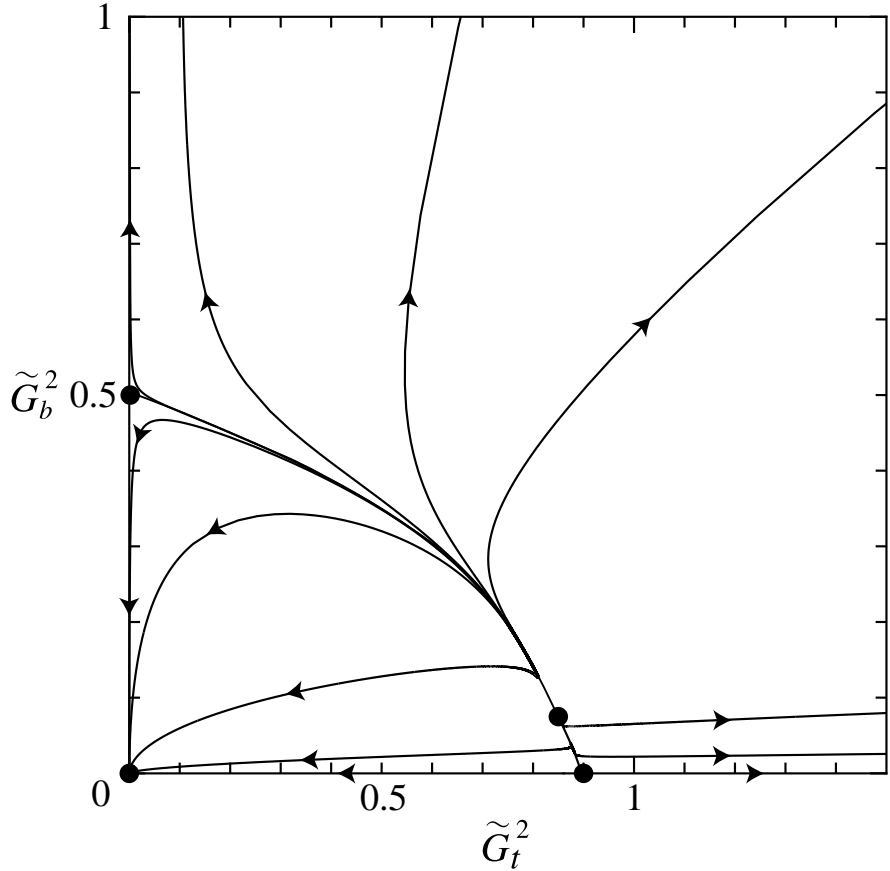}}
\caption{RG flows in the space of 
$\tilde{G}_t^2$ and $\tilde{G}_b^2$.
The fixed points are denoted by a bullet.}
        \label{fig:3}
        \end{figure}
\begin{figure}
\epsfxsize= 12 cm
\centerline{\epsffile{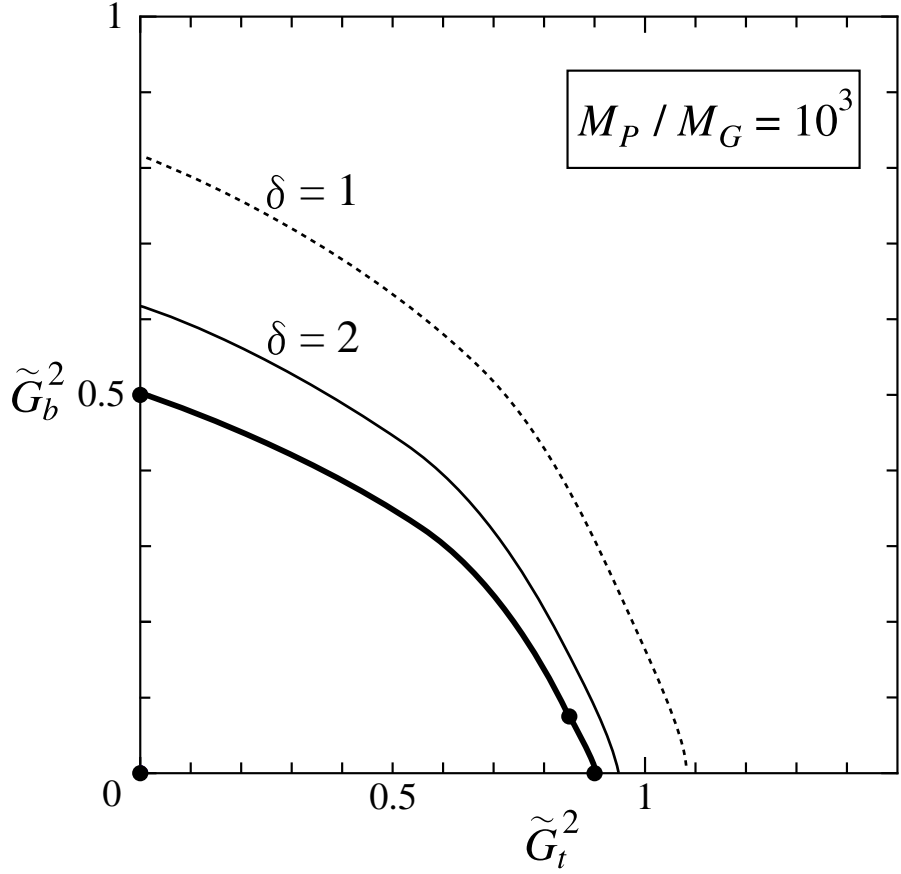}}
\caption{ The stable manifold (bounded by the bold line)
 in the space of 
$\tilde{G}_t^2$ and $\tilde{G}_b^2$.
The regions extended by the relaxed nontriviality requirement
are also shown; the dotted,   straight boundary lines
correspond to $\delta=1,2$, respectively. As for $\delta=0$,
the whole region in Fig. 4 satisfies the relaxed nontriviality
requirement.}
        \label{fig:4}
        \end{figure}

The nontriviality requirement above could be to strong;
it is a requirement in the  $\Lambda \to \infty$ limit.
It can be relaxed so as to require
for the couplings not to develop into a Landau pole
before the Planck scale $M_P$. 
Since the above result on the fixed points (\ref{fps})
is independent on the energy scale $\Lambda$,  especially on
the actual value of the unification scale $M_G$,
the ratio $M_P/M_G$ can take any value greater  than  $ \sim 10^3$.
Clearly, the smaller the ratio is, the milder is the 
relaxed nontriviality requirement. In Fig. 4 we show how the 
allowed region is extended by 
relaxing of the nontriviality requirement in the case of $M_P/M_G= 10^3$.
The relaxed condition depends on
the number of the extra dimensions $\delta$. We have
considered three cases $\delta=0, 1$ and $2$ in Fig. 4,
where the stable manifold is bounded by the bold line.
As we can see from Fig. 4, the extensions
for $\delta=1$ and $2$ are very small. 
This is a consequence of the power law running of the couplings;
the couplings evolve faster in extra dimensions as the energy scale
varies, and so the Landau pole can be faster reached compared with
the case of the logarithmic running. Therefore, the initial point
cannot be very far from the stable manifold.
As for   the logarithmic 
running ($\delta=0$)  \footnote{Here we are
interested only in the qualitative  nature. 
So, to derive the allowed region in
the case  of the logarithmic running, 
we have used the RG equations 
(\ref{beta-t}) and  (\ref{beta-b})
for $\tilde{G}_t$ and $\tilde{G}_b$, 
while for the gauge coupling we have used Eq. (\ref{beta-ggut}).
with $X_\delta (R \Lambda)^\delta=1$.}, we found that the whole region of Fig. 4
satisfies the relaxed nontriviality
requirement ($M_P/M_G= 10^3$), justifying our statement above.
In the next subsection we would like to investigate
phenomenological consequences from the nontriviality requirement.

\subsection{The model between $1/R$ and $M_G$}
To be more specific we assume that the extra dimensions
are large, i.e. $1/R < 1/10^{16}$ GeV, and that for energies
below $1/R$ the MSSM is the effective theory.
For the energy scale between $1/R$ and $M_G$,
the effective theory is exactly the 
one proposed in Ref. \cite{dienes1}, in which
only the gauge boson and Higgs supermultiplets of the MSSM
have a tower of Kaluza-Klein states and 
the lepton and quark supermultiplets have no  tower of
Kaluza-Klein states. Correspondingly,
the one-loop $\beta$-functions for the energy scales
between $1/R$ and $M_G$
become \cite{dienes1}
\be
16 \pi^2\beta_1&=&
 g_1^3~[6+\frac{6}{5}(X_\delta/2)
(\Lambda R)^{\delta}]~, \\
16 \pi^2\beta_2&=&
g_2^3~[4-6(X_\delta/2) (\Lambda R)^{\delta}]~, \\
16 \pi^2\beta_3&=&
g_3^3~[3-12(X_\delta/2) (\Lambda R)^{\delta}]~,
\label{betas}\\
16 \pi^2\beta_t&=&
G_t~[3 G_t^2-\frac{3}{10}g_1^2-\frac{3}{2} g_2^2 \\
& &+(X_\delta/2)~ (\Lambda R)^{\delta}
(6G_t^2+2G_b^2 -\frac{17}{15}
g_1^2-3g_2^2-\frac{32}{3}g_3^2)]~, \\
16 \pi^2\beta_b&=&
G_b~[3 G_b^2+G_\tau^2-\frac{3}{10}g_1^2-\frac{3}{2} g_2^2 \\
& &+ (X_\delta/2)~ (\Lambda R)^{\delta}
(2G_t^2+6G_b^2 -\frac{1}{3}g_1^2-3g_2^2-\frac{32}{3}g_3^2)]~, \\
16 \pi^2\beta_\tau&=&
G_\tau~[3 G_b^2+G_\tau^2-\frac{3}{10}g_1^2-\frac{3}{2}g_2^2 \\
& &+ (X_\delta/2) ~(\Lambda R)^{\delta}
(6G_\tau^2 -3g_1^2-3g_2^2)]~, 
\ee
where $g_{1,2,3}$ are the gauge couplings, 
$G_{t,b,\tau}$ are the Yukawa
couplings for the top, bottom and tau, in the MSSM, respectively.
We have neglected other Yukawa couplings, and
use has been made of the fact that the
anomalous dimensions  of the Higgs supermultiplets due
to $N=2$  supersymmetry in the excited sector vanish between
$1/R$ and $M_G$ \cite{howe}.

\subsection{The lower bound of $\tan\beta$}

In what follows we will consider only 
the case with $\delta =1$. Moreover, to simplify the situation, we assume  
that there exists  a uniform
SUSY threshold $M_{SUSY}$.
We study the evolution
of the couplings
below $1/R$  at the two-loop level
\footnote{See Ref. \cite{kubo1} for more
details of the method of the present analyses.}, along
with the experimental inputs \cite{pdg}; the tau mass
$ M_\tau=1.777 ~{\rm GeV}$, 
the $Z$ gauge boson mass $M_Z=91.187 ~{\rm GeV}$,
the effective electromagnetic coupling 
$\alpha^{-1}_{\rm EM}(M_Z) = 127.9$ at $M_Z$, and
the Weinberg mixing angle
$\sin ^2 \theta_W(M_Z)=0.2312$
in the modified minimal subtraction scheme. The experimental value of 
the physical top mass is given by \cite{pdg} 
\be
M_t &=& (174.3\pm 5.1) ~~\mbox{GeV}~.
\label{mt}
\ee

At the SUSY threshold $M_{SUSY}$ we require that the matching conditions, 
\be
G_{t}^{\rm SM} 
&=&G_{t}\,\sin \beta~,~
G_{b}^{\rm SM}
~ =~ G_{b}\,\cos \beta~,
~G_{\tau}^{\rm SM}
~=~G_{\tau}\,\cos \beta~,\nn\\
\lambda &=&
\frac{1}{4}(\frac{3}{5}g^2_{1}
+g^2_2)\,\cos^2 2\beta~,
\ee
should be satisfied,
where $G_{i}^{\rm SM}~(i=t,b,\tau)$ are
the SM Yukawa couplings and $\lambda$ is the Higgs self-coupling.
This is our definition of $\tan\beta$.
 (There are MSSM threshold corrections
to this matching condition \cite{hall2,wright1}, which
we ignore in the following discussion.)
For a given set of the initial values of $G_t$ and $G_b$ at $M_{G}$,
the top quark mass $M_t$ is no longer a free parameter and can be computed, where we
use the formula \cite{barger,hall2}
\be
M_{t} &=&m_{t}(M_t)\,[\,1+
\frac{4}{3}\frac{\alpha_3(M_t)}{\pi}+
10.95\,(\frac{\alpha_3(M_t)}{\pi})^2-0.3
\frac{\alpha_t(M_t)}{\pi}\,]~.
\label{topmass}
\ee
Here $\alpha_3=g_3^2/4\pi$, $\alpha_t=(G_t^{SM})^2/4\pi$, and
$m_{t}(\mu)$ is the running top mass in the
modified minimal subtraction scheme and given by
\be
m_t(\mu)&=&\frac{1}{\sqrt{2}}G_t^{MS}(\mu)\,v(\mu)~
\mbox{with} ~~v(M_Z)=246.22~\mbox{GeV}~,
\ee
where $v$ is the vacuum expectation value
of the SM Higgs field which is made of the two Higgs fields
of the MSSM. 
The mass of the bottom quark
can suffer from a large correction from
the SSB terms \cite{hall2,wright1}. But 
we do not take into account them in the present analysis,
because we do not consider the SSB terms.

Given all the facilities for the RG evolution of the 
couplings, we 
choose a value for $\tan\beta$ with the top mass
varying from $170$ to $180$ GeV  and
let evolve the couplings from  $M_Z$ to $M_G$
(at which the gauge coupling unification and the $b-\tau$ unification
is realized). We then calculate the ratio
\be
k_t &=&\frac{G_t^2}{g_G^2}=\tilde{G}_t^2
\ee
at $M_G$ as a function
of $\tan\beta$, where $g_G$ is the unified gauge coupling.
The results are shown in Figs. 5, 6 and 7.
In Fig. 5 we vary $\tan\beta$ from $1$ to $60$ with
$M_{SUSY}=1$ TeV, where the straight (dotted) line stands
for $1/R=10^{14} (10^{5})$ GeV.
The range of smaller $\tan\beta$ are plotted in Fig. 6 and 7;
Fig. 6 shows the $M_{SUSY}$ dependence for $1/R$ fixed at $ 10^{14} $ GeV,
while 
Fig. 7 shows the $R$ dependence for $M_{SUSY}$ fixed at $1$ GeV.
We  see from these
figures that the value of $k_t$ increases rapidly as $\tan\beta$
approaches $\sim 2$ from larger values, and
that this feature does not depend very much on
$R$ and $M_{SUSY}$.
Comparing this result with Fig. 4 (which
shows the region in the $\tilde{G}_t^2$-$\tilde{G}_b^2$ plane
satisfying the nontriviality requirement),
we see that 
for a small value ($\lsim 2$) of $\tan\beta$,
leading to a large value of $k_t$,
the theory cannot be made nontrivial. 

As we have seen in Sect.4.2, the difference between the power law and
logarithmic running is 
how fast the RG evolution develops into
a Landau pole as $\Lambda$ increases.
Moreover, the more there exist extra dimensions, the faster is the evolution,
and hence the closer to
the stable manifold is
the region satisfying
the relaxed nontriviality requirement (see in Fig. 4).
From this observation we  conclude that the presence of
the extra dimensions prefers a large value ($\gsim 2$) of $\tan\beta$.
As it is known \cite{higgsmass}, the mass
of the MSSM Higgs depends on $\tan\beta$. 
The search for the Higgs particle at LEP2 have already 
excluded the range of $tan\beta$ \cite{pdg,igo}
\be
\left\{\begin{array}{l}
 0.5 \sim 2.3 \\
0.7 \sim 1.9 
\end{array}\right.
~~\mbox{for} ~~M_t = 
\left\{ \begin{array}{l}
175 \\
180
\end{array}\right.~\mbox{GeV}~.
\ee
So today's possible observation of the Higgs particle \cite{lep2}
might be an indication of the existence of extra dimensions.
\begin{figure}
\epsfxsize= 12 cm
\centerline{\epsffile{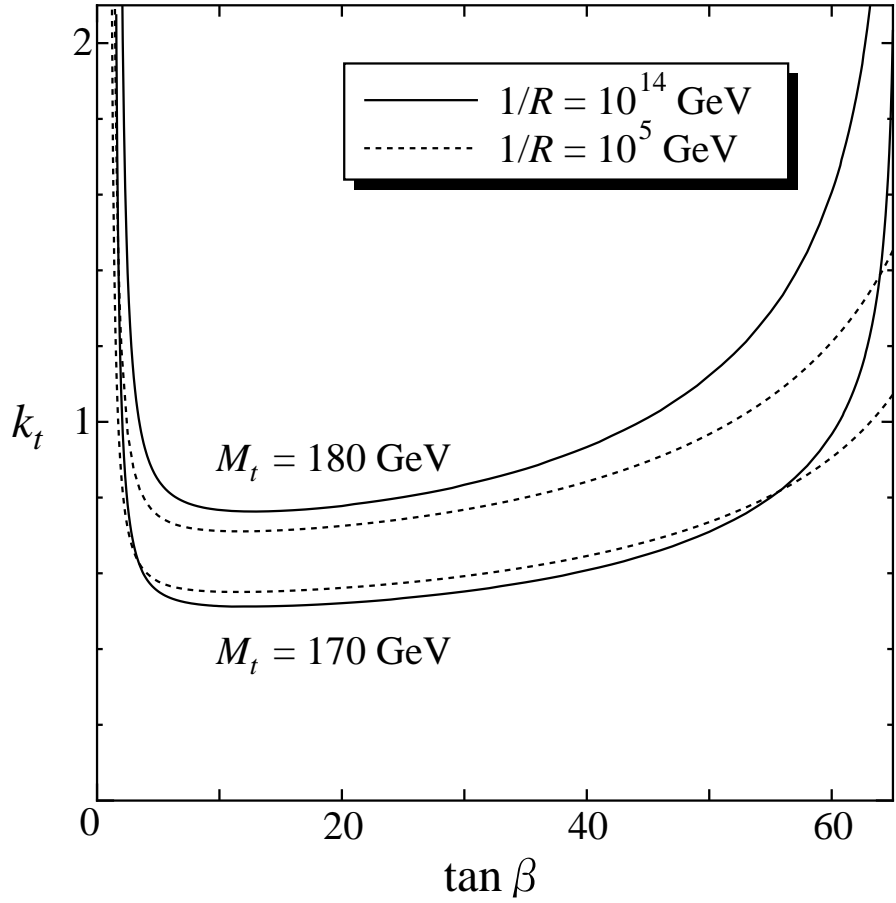}}
\caption{$k_t$ as a function of $\tan\beta$
for $1/R = 10^{14}$ (straight) and $10^5$ (dotted) GeV
with $M_{SUSY}=1$ TeV.}
        \label{fig:5}
        \end{figure}

\begin{figure}
\epsfxsize= 12 cm
\centerline{\epsffile{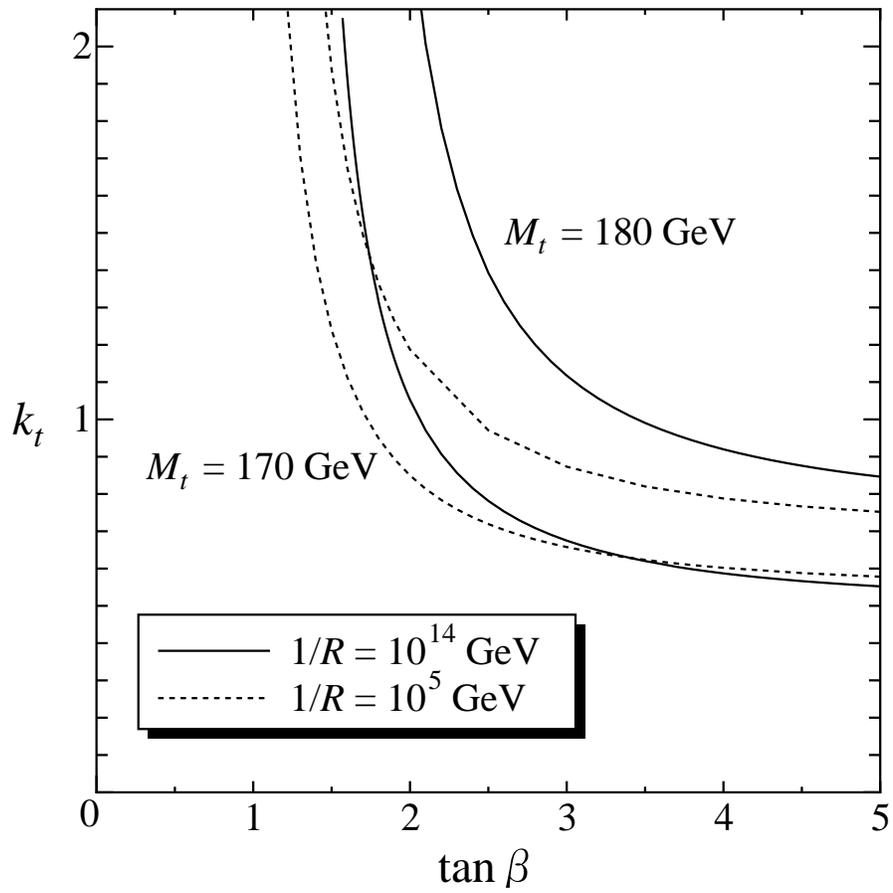}}
\caption{ The same as Fig. 5
near $\tan\beta=2$.}
        \label{fig:6}
        \end{figure}

\begin{figure}
\epsfxsize= 12 cm
\centerline{\epsffile{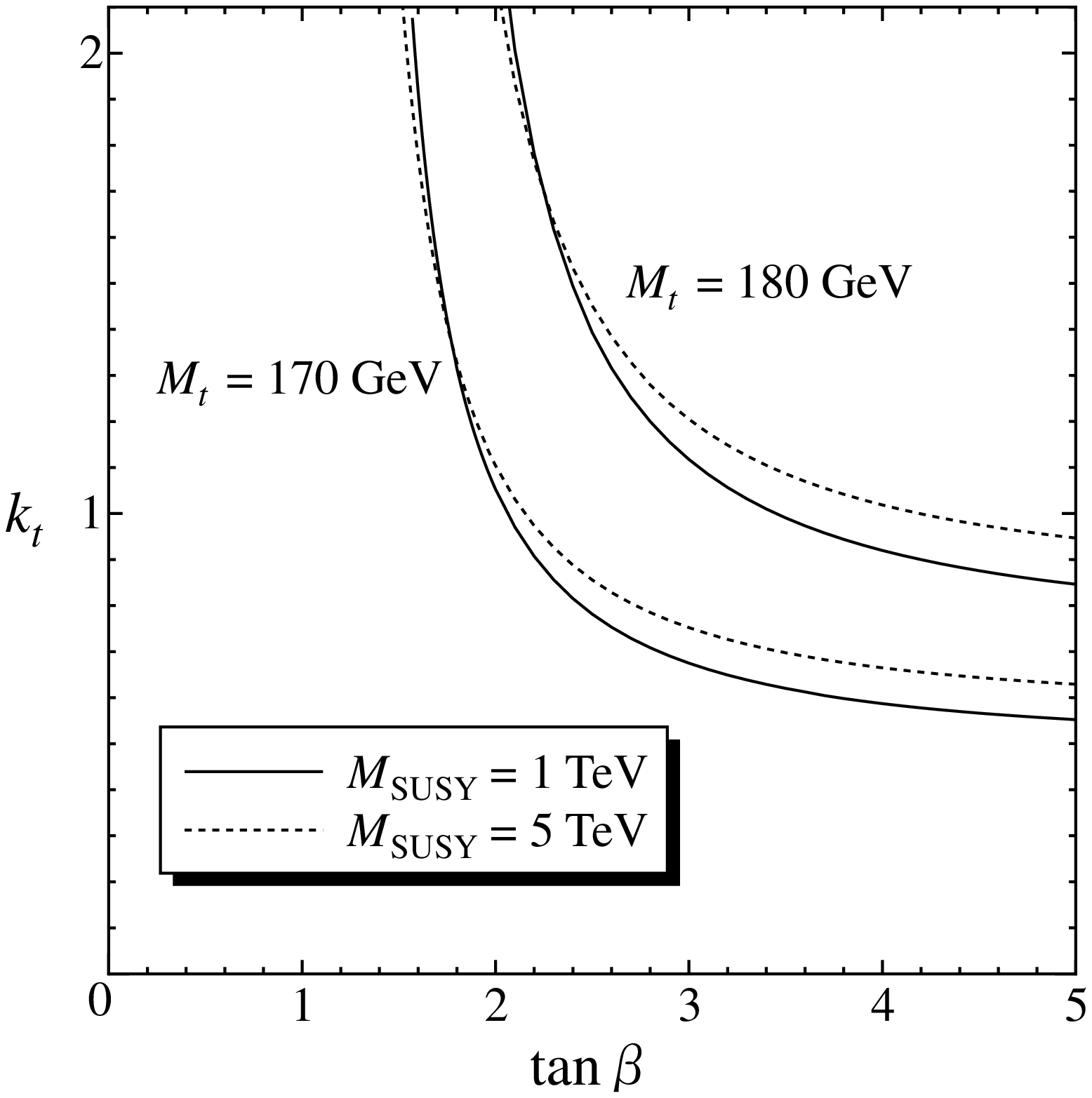}}
\caption{The $M_{SUSY}$
dependence of $k_t$ near $\tan\beta=2$ with
 $1/R = 10^{14}$ GeV.}
        \label{fig:7}
        \end{figure}

\section{Conclusion}
Our starting point was to assume
that the result obtained from the 
Monte Carlo simulations on
the nonperturbative existence of
the Yang-Mills theory in five dimensions \cite{ejiri1}
can be  applied to a more general class of higher-dimensional 
unified gauge theories.
The first nontrivial requirement is that the theory should be
in phase I of Fig. 1, because otherwise 
the massive Kaluza-Klein excitations would not decouple
at low energies. 
Then we  have derived  the conditions (\ref{cond1}) imposed by
the nontriviality requirement
on the supersymmetric gauge theories
containing matter superfields,
where we have also  considered
relaxing the nontriviality requirement.
These results have been applied  to
a concrete SUSY GUT based on $SU(5)$, 
and we have found, comparing Fig. 4 with Figs. 5, 6 and 7  that 
the model prefers a large value 
($\gsim 2$) of $\tan\beta$. Moreover it
has been argued that this is not a model-specific feature,
 but  a general feature of 
SUSY GUTs with extra dimensions.

\noindent
{\large \bf Acknowledgments}\\
This work is supported partially by  the Ministry of
Education, Science 
and Culture and by
 the Japan Society for the Promotion of Science.
We would like to thank H. Nakano, D. Suematsu  and H. Terao 
for useful discussions.

\end{document}